\documentclass[lettersize,journal]{IEEEtran}
\usepackage{amsmath,amsfonts}
\usepackage{algorithmic}
\usepackage{array}
\usepackage[caption=false,font=normalsize,labelfont=sf,textfont=sf]{subfig}
\usepackage{textcomp}
\usepackage{stfloats}
\usepackage{url}
\usepackage{verbatim}
\usepackage{graphicx}
\def\BibTeX{{\rm B\kern-.05em{\sc i\kern-.025em b}\kern-.08em
    T\kern-.1667em\lower.7ex\hbox{E}\kern-.125emX}}
\usepackage{balance}
\usepackage{booktabs}
\usepackage{multirow}

\begin{document}
\title{GC360IQ: Generic-to-Individualized Quality Assessment for Stitched 360-Degree Panoramas}
\author{Jinghan~Zhou,~\IEEEmembership{Student~Member,~IEEE}, 
        Zhou~Wang,~\IEEEmembership{Fellow,~IEEE}%
\thanks{Both authors are with the Department of Electrical and Computer Engineering, University of Waterloo, ON, Canada.}
\thanks{This work has been submitted to the IEEE for possible publication.
Copyright may be transferred without notice, after which this version may no longer be accessible.}}


\maketitle

\begin{abstract}
Existing image quality assessment (IQA) methods typically predict mean opinion scores (MOSs) but struggle to capture variations in individual subject behaviors. This limitation is highly pronounced in immersive visual applications such as stitched 360-degree panoramas, where user opinions diverge drastically based on personal sensitivity to blending-induced luminance inconsistency, detail loss, and geometric misalignment. Here we propose GC360IQ, a novel Generic-to-Individualized IQA framework that establishes a learned explicit feature space to characterize individual subject behaviors. First, we construct a specialized 360-degree panorama database focusing on blending-induced luminance and detail degradation while minimizing geometric misalignment, providing multidimensional quality ratings alongside complete individual scores. Second, we develop a generic quality model that utilizes unblended views as a perceptual reference. Dual feature extraction branches capture gradient and structural information specifically around stitching regions to predict baseline quality. Third, we construct a compact preference embedding space that acts as an explicit feature domain to model each subject's deviation from generic quality perceptions. We also introduce a maximum a posteriori (MAP) adaptation mechanism. By leveraging a preference prior learned within our explicit feature space, a new subject's unique behavioral embedding is mapped progressively with an increasing number of anchor ratings. Experiments demonstrate that the generic model provides accurate stitching quality predictions and that subject adaptation further improves individual score predictions. Deeper analysis of the collected ratings and learned embeddings reveals that observer differences contain structured variation related to scoring tendencies and sensitivity to stitching artifacts, rather than merely random rating noise.
\end{abstract}


\maketitle
\section{Introduction}

Most existing image quality assessment (IQA) models are trained to predict the mean opinion score (MOS), which summarizes subjective ratings into a single measure of perceived quality. However, this averaging process implicitly treats observers as a homogeneous group and may conceal consistent differences in how individuals judge the same image. Streijl et al.~\cite{streijl2016mean} discussed the limitations of MOS in representing observer variability, while Engelke et al.~\cite{engelke2012human} showed that inter-observer differences contain perceptually meaningful information. More recent studies have further reported stable individual visual preferences~\cite{cherepkova2024individual}, suggesting that observer disagreement may reflect systematic perceptual variation rather than only random rating noise. These observations motivate individualized IQA models that preserve such observer differences rather than treating them only as variation around MOS.

A conventional MOS-oriented model can provide a generic quality estimate across observers, but it cannot describe how a particular observer consistently deviates from that estimate. Individualized quality assessment therefore requires a representation that captures subject-specific scoring tendencies and sensitivities. An important challenge is to construct an explicit learned preference space that characterizes these behavioral differences and to estimate the preference embedding of a new subject from only a limited number of subject-specific ratings.

This limitation is particularly pronounced in immersive visual applications such as stitched 360-degree panoramas. Stitched panoramas are widely used to create immersive visual experiences in virtual reality and are generated by combining multiple camera views into a single spherical image, where imperfect alignment and blending may introduce geometric misalignment, luminance inconsistency, and detail loss. Recent perceptual psychology research suggests that immersive visual experience depends not only on sensory fidelity but also on observers' expectations of realism and motion coherence~\cite{murovec2024role}. In stitched panoramas, luminance inconsistency and detail loss are among the most influential artifacts, arising mainly from insufficient exposure compensation and excessive smoothing during blending~\cite{yang2017content,madhusudana2019subjective}. Individual observers may attend to, tolerate, or weight these localized artifacts differently, leading to distinct quality judgments that are not fully represented by their average.

Accurate individualized prediction in this setting first requires a reliable generic quality estimate. However, the perceptual quality of stitched panoramas is difficult to assess because no pixel-aligned pristine panorama is available, while stitching introduces spatially localized distortions around seam regions. These distortions differ from conventional degradations such as compression, blur, and projection artifacts commonly considered in image quality assessment. Together, these observations support a perception-referenced view of stitched panorama quality, in which human judgments are guided by expected luminance continuity and structural fidelity rather than direct pixel-level correspondence to a pristine reference.

To address these challenges, we present \textbf{GC360IQ}, a \textbf{Generic-to-Individualized} quality assessment framework for stitched 360-degree panoramas. We construct a dedicated database focusing on blending-induced luminance inconsistency and detail loss while minimizing geometric misalignment. The database provides multidimensional quality annotations together with complete subject-level ratings. The generic model uses the unblended views as a perceptual reference and learns artifact-aware image representations by extracting gradient and structural information around stitching regions to predict baseline quality. The individualized model constructs a compact preference embedding space that explicitly represents each subject's deviation from the generic quality prediction. By leveraging the learned preference prior, a new subject is progressively mapped into the preference embedding space as more anchor ratings become available through maximum a posteriori (MAP) adaptation. A preliminary study of the GC360IQ database and generic quality assessment was presented in our previous work~\cite{zhou2025gc360iq}.

The main contributions of this work are summarized as follows:

\begin{itemize}

\item We construct a dedicated subject-level quality assessment database for stitched 360-degree panoramas, providing multidimensional quality annotations and complete individual ratings for the study of generic and individualized quality perception.

\item We propose a perception-referenced Generic-to-Individualized quality assessment framework that combines artifact-aware generic quality modeling with a compact preference embedding space and MAP-based subject adaptation from limited anchor ratings.

\item We reveal that subject-level deviations from MOS contain structured low-dimensional variation and show that individualized adaptation is particularly beneficial for observers poorly represented by the average score, preserving consistent perceptual differences rather than treating them as random rating noise.

\end{itemize}

\section{Related Work}
\label{sec:related_work}

\subsection{360-Degree Image Quality Databases}

A number of subjective databases have been developed for omnidirectional image quality assessment. 
Early databases mainly considered spatially homogeneous distortions, such as compression, blur, noise, resolution reduction, and projection artifacts. 
Representative examples include Upenik2016/2017~\cite{Upenik2016,Upenik2017}, CVIQD~\cite{cviqd}, CVIQD2018~\cite{cviqd2018}, the database of Huang \textit{et al.}~\cite{huang2018}, and OIQA~\cite{oiqa2018}. 
These databases have provided important benchmarks for general-purpose 360-degree IQA.

Recent studies have extended OIQA databases to spatially varying distortions.
JUFE~\cite{liu2024perceptual} studies locally distorted omnidirectional images under different viewing conditions. 
OIQ-10K~\cite{yan2025omnidirectional} covers both homogeneous and heterogeneous distortions and provides annotations for perceived quality, distortion regions, head movement, and eye movement. 
Yan \textit{et al.}~\cite{yan2025subjective} further constructed a large database in which distortions are introduced into one or two camera views. 
These studies show that both distortion distribution and viewing behavior affect the perceived quality of omnidirectional images.

Several databases focus specifically on panorama stitching. 
Yang \textit{et al.}~\cite{yang2017content} constructed a synthetic stitched image database, ISIQA~\cite{madhusudana2019subjective} includes multiple artifacts arising from alignment and panorama generation, and CROSS~\cite{li2019cross} contains panoramas produced by different stitching algorithms.
Existing stitched image databases provide limited support for isolating blending-induced luminance inconsistency and detail loss from geometric stitching artifacts.

\subsection{360-Degree Image Quality Assessment}

Conventional full-reference and no-reference IQA methods, including SSIM~\cite{wang2004imageSSIM}, GMSD~\cite{xue2013gradient}, BRISQUE~\cite{mittal2012noBrisque}, and NIQE~\cite{mittal2012makingNIQE}, have been widely used to assess common image distortions. 
Building on conventional metrics, early omnidirectional IQA methods adapted quality computation to spherical viewing geometry. 
Representative methods such as WS-PSNR~\cite{xiu2017evaluationAW-SPSNR} and WS-SSIM~\cite{zhou2018weightedWS-SSIM} introduce spherical weighting to compensate for the nonuniform sampling of equirectangular projections.

Recent NR methods based on deep learning have moved from direct analysis in the projection domain to quality modeling with viewports. 
VGCN~\cite{xu2020blind} uses graph convolution to model local viewport quality and relations among viewports, while ST360IQ~\cite{tofighi2023st360iq} uses saliency to guide tangent viewport sampling and applies a spherical vision transformer. 
More recent studies have considered viewing behavior and spatially varying distortions. 
Assessor360~\cite{wu2023assessor360} generates multiple viewport sequences to approximate how different observers browse an omnidirectional image. 
Max360IQ~\cite{yan2025max360iq} uses attention along multiple axes and features at several scales to capture global degradation and localized distortion in both uniformly and spatially nonuniformly distorted panoramas. 
IQCaption360~\cite{yan2025omnidirectional} jointly predicts distortion distribution and perceptual quality, producing a quality score and a quality description generated from a predefined template.

Only a few studies have addressed stitched image quality directly.
Yang \textit{et al.}~\cite{yang2017content} studied synthetic stitched panoramas and proposed content-aware quality evaluation for stitching distortions. 
Madhusudana \textit{et al.}~\cite{madhusudana2019subjective} investigated objective assessment across different stitching pipelines and artifacts. 
These studies identify geometric misalignment, ghosting, luminance inconsistency, and structural degradation as important factors in stitched-image quality. 
Existing methods remain centered on aggregate stitching quality and do not predict individual responses to blending artifacts.

\subsection{Subject Variation and Individualized Visual Assessment}
\label{sec:review_sub_depend}

Most IQA studies use the mean opinion score (MOS) as the subjective ground truth. 
However, MOS may conceal meaningful differences among observers~\cite{streijl2016mean,engelke2012human}. 
Recent evidence further suggests that individual visual preferences can remain stable across repeated observations~\cite{cherepkova2024individual}.
Statistical models have also been used to analyze individual opinion scores. 
Li and Bampis~\cite{li2017recover} jointly estimated latent stimulus quality, observer bias and consistency, and content ambiguity from noisy ratings. 
Li \emph{et al.}~\cite{li2020simple} later modeled observer behavior using bias and inconsistency, producing an aggregate quality estimate corrected for bias and weighted by consistency. 
These methods mainly use individual ratings to improve aggregate quality estimates rather than to predict each observer's judgment.

Individualized visual assessment instead models variation among observers as part
of the prediction target. Yang et al.~\cite{yang2022personalized} incorporated
subject attributes into aesthetic assessment, while studies across age
groups~\cite{wang2023age} found systematic differences in perceived image
quality. Individualized quality-of-experience models~\cite{huang2022personalized,
jia2025towards} have similarly used user characteristics to improve perceptual
prediction and system optimization.

Despite this progress, individualized quality prediction for
blending artifacts in stitched 360-degree panoramas has received
little attention.

\section{Proposed Method}
\label{sec:method}

\begin{figure*}[t]
    \centering
    \includegraphics[width=0.98\textwidth]{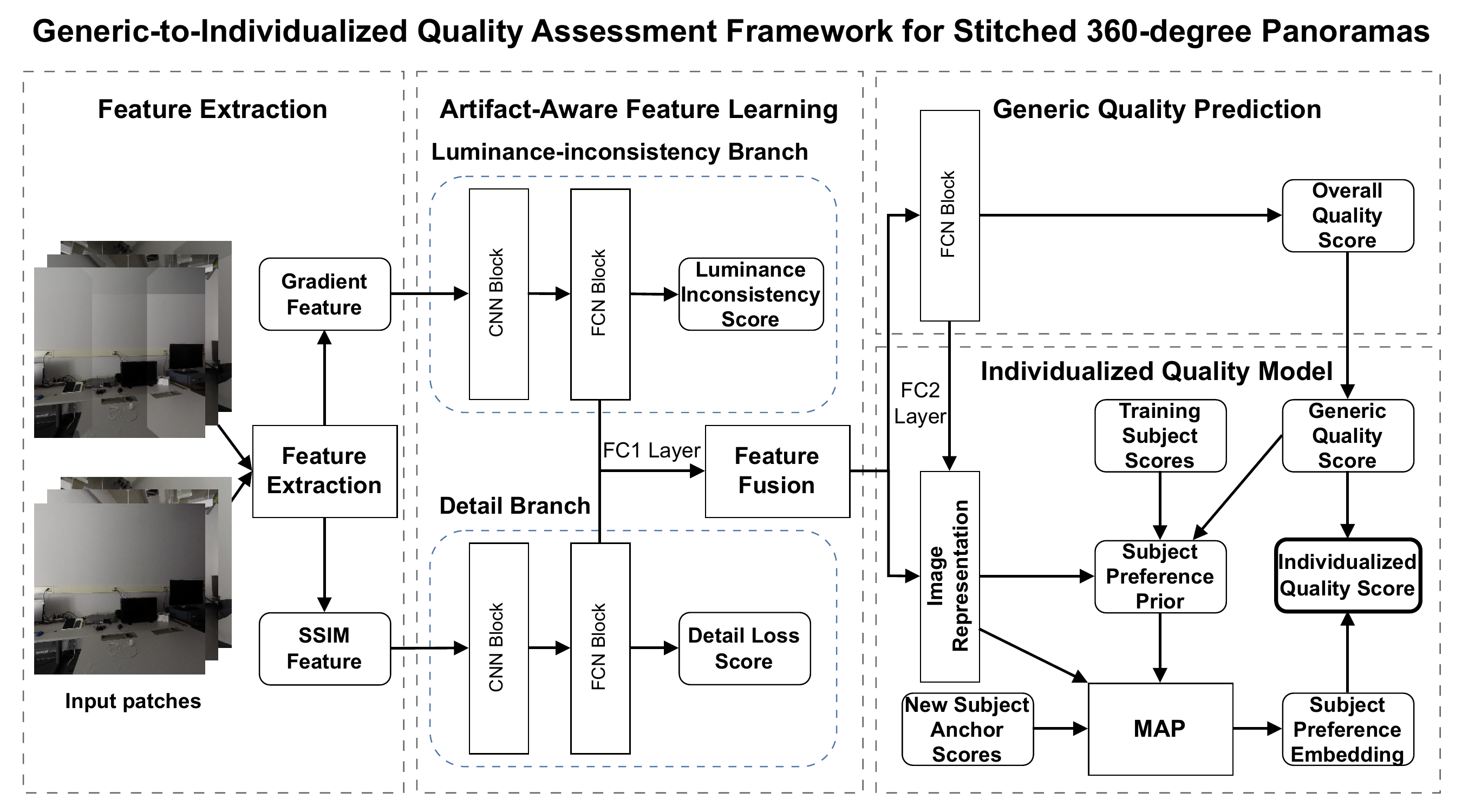}
    \caption{
    Overview of the proposed GC360IQ framework for generic and individualized quality assessment.
    The generic model learns representations of luminance inconsistency and detail loss, while the individualized model adapts the generic prediction to each subject using limited ratings.
    }
    \label{fig:overall_model}
\end{figure*}

This section presents the proposed GC360IQ framework for generic and individualized quality prediction. 
As shown in Fig.~\ref{fig:overall_model}, local viewports are extracted around the stitching regions from both the stitched panorama and its unblended inputs. 
The generic model evaluates luminance inconsistency and detail loss with two separate branches, and combines their features to predict overall quality. 
These features are also used for individualized prediction. 
A preference prior is first learned from the training subjects. 
Given a small number of anchor ratings from a new subject, maximum a posteriori (MAP) adaptation estimates the corresponding preference embedding. 
The estimated embedding is then used to adjust the generic prediction and predict the subject's scores for the remaining images.
The following subsections describe the preprocessing pipeline, the generic quality model, and the individualized quality model.

\subsection{Preprocessing}
\label{sec:preprocessing}

To provide spatially aligned inputs for artifact-aware quality modeling, 
the unblended input images and the stitched panorama are processed jointly for each scenario. 
Following the equirectangular FOV extraction strategy~\cite{Lee2021Viewport}, 
each panorama is projected into seven viewport directions. 
Perspective views are cropped from these directions and resized to $256\times256$ pixels. 
A seam-boundary mask is generated using the same projection to define valid seam regions.

Each preprocessed sample is represented as $(\mathbf{x}, \mathbf{y}, \mathbf{p})$, 
where $\mathbf{x}$ and $\mathbf{y}$ denote the stitched and unblended patch stacks, respectively, 
and $\mathbf{p}$ is a binary seam mask. 
The mask is applied to the feature maps in both artifact branches to retain seam regions and reduce the influence of non-overlapping areas.

\subsection{Generic Quality Model}
\label{sec:gen_model}

The generic quality model predicts overall quality using MOS as the training target and provides image representations for subject adaptation. 
It evaluates luminance inconsistency and detail loss in two separate branches, whose features are then fused to produce the overall quality score.

Luminance inconsistency between the stitched and unblended patches 
is estimated from spatial gradients. 
The patches are convolved with the $11{\times}11$ horizontal and vertical derivative filters 
$\mathbf{h}=\mathbf{u_0}\mathbf{u_1}^\top$ and 
$\mathbf{v}=\mathbf{u_1}\mathbf{u_0}^\top$ following the Farid--Simoncelli design~\cite{farid2004differentiation}. 
The resulting gradient matrices are 
$\mathbf{S}=[\mathbf{x}*\mathbf{h},\;\mathbf{x}*\mathbf{v}]$ and 
$\mathbf{O}=[\mathbf{y}*\mathbf{h},\;\mathbf{y}*\mathbf{v}]$, 
where $*$ denotes 2D convolution with symmetric padding to preserve boundary continuity~\cite{wang2004imageSSIM}. 

Luminance inconsistency is characterized by the gradient magnitude and directional differences:
\begin{equation}
a_{m,n}=\lVert\mathbf{s}_{m,n}-\mathbf{o}_{m,n}\rVert_2, \quad
b_{m,n}=1-\frac{\mathbf{s}_{m,n}\cdot\mathbf{o}_{m,n}}
{\lVert\mathbf{s}_{m,n}\rVert_2\lVert\mathbf{o}_{m,n}\rVert_2},
\label{eq:luminance_diff}
\end{equation}
where $\mathbf{s}_{m,n}$ and $\mathbf{o}_{m,n}$ are local gradient vectors at pixel $(m,n)$.
Seven pairs of $\mathbf{A}=\{a_{m,n}\}$ and $\mathbf{B}=\{b_{m,n}\}$ maps from different FOVs are concatenated as 
$[\mathbf{A}_1,\mathbf{B}_1,\;\dots,\;\mathbf{A}_7,\mathbf{B}_7]$
to form a 14-channel tensor $\boldsymbol{\eta}\in\mathbb{R}^{14\times256\times256}$, 
which is input to a CNN $f_{\text{CNN}}$ for luminance-inconsistency score prediction. 
The CNN comprises five convolutional and two fully connected layers, 
with feature depth expanding from 14 to 64 channels via alternating $5\times5$ and $3\times3$ kernels 
(strides $\{2,1,2,1,2\}$), each followed by ReLU activation and $2\times2$ max pooling, 
except for the final linear output layer.

Detail loss is reflected by luminance, contrast, and structural variations between the stitched and unblended images. 
Following the SSIM formulation~\cite{wang2004imageSSIM}, the luminance term, contrast term, and full SSIM response are computed within an $11\times11$ sliding window using symmetric padding. 
For each of the seven FOVs, the luminance maps $\mathbf{L}$, contrast maps $\mathbf{C}$, and SSIM maps $\mathbf{D}$ are obtained and concatenated in the same manner as in the luminance-inconsistency branch:
$[\mathbf{L}_1,\mathbf{C}_1,\mathbf{D}_1,\;\dots,\;\mathbf{L}_7,\mathbf{C}_7,\mathbf{D}_7].$
This forms a 21-channel tensor
$\boldsymbol{\zeta}\in\mathbb{R}^{21\times256\times256}$,
which is fed into the CNN $g_{\mathrm{CNN}}$ for detail-loss score prediction.
The network shares the same architecture as $f_{\mathrm{CNN}}$, except that its input contains 21 channels instead of 14 to incorporate the detail-related feature maps.

The generic overall quality representation is obtained by fusing intermediate features from the two artifact branches.  
Feature vectors from the first fully connected layers, 
$\mathbf{f}_{\text{fc1}}$ and $\mathbf{g}_{\text{fc1}}$, 
are concatenated as $[\mathbf{f}_{\text{fc1}},\mathbf{g}_{\text{fc1}}]$ 
to form a 128-dimensional feature $\boldsymbol{\Phi}\in\mathbb{R}^{128}$.  
A three-layer fully connected network $t_{\text{FCN}}$ 
with hidden dimensions 128 and 256 and ReLU activations 
produces the generic overall quality prediction:
\begin{equation}
Q_i^{\mathrm{gen}} = t_{\text{FCN}}(\boldsymbol{\Phi}_i).
\end{equation}
The intermediate features of the artifact branches and the quality-fusion network are further used to construct the image-level representation for subject adaptation in Sec.~\ref{sec:ind_model}.

\subsection{Individualized Quality Model}
\label{sec:ind_model}

The individualized quality model represents each subject's quality score as a generic prediction plus a residual. 
The residual is parameterized by a preference embedding estimated from a small set of anchor ratings with a prior learned from the training subjects. 
For an image $i$ and subject $s$, the quality score is modeled as
\begin{equation}
Q_{i,s}
=
Q_i^{\mathrm{gen}}
+
r_{i,s},
\label{eq:subject_residual}
\end{equation}
where $Q_i^{\mathrm{gen}}$ denotes the generic quality prediction and $r_{i,s}$ is the deviation of subject $s$ from this prediction.

To characterize such residuals, we construct an image-level representation from the intermediate features of the generic quality model. 
Specifically, the first fully connected features from the luminance-inconsistency and detail-loss branches, 
$\mathbf{f}_{\mathrm{fc1},i}$ and $\mathbf{g}_{\mathrm{fc1},i}$, 
are concatenated with the second fully connected feature from the overall quality fusion network, 
$\mathbf{t}_{\mathrm{fc2},i}$, to form
$
\boldsymbol{\Psi}_i
=
[
\mathbf{f}_{\mathrm{fc1},i},
\mathbf{g}_{\mathrm{fc1},i},
\mathbf{t}_{\mathrm{fc2},i}
].
$
This representation combines artifact-specific cues with the higher-level generic quality representation. 
It is then reduced to a 64-dimensional vector using PCA and augmented with a constant bias dimension, resulting in 
$\mathbf{z}_i \in \mathbb{R}^{D+1}$, where $D=64$. The PCA projection is estimated using only the training images and remains fixed during testing. The PCA features are further standardized to zero mean and unit variance before preference estimation.
The residual is then approximated by a linear subject-dependent model:
\begin{equation}
r_{i,s}
=
\mathbf{z}_i^{\top}
\boldsymbol{\theta}_s,
\label{eq:linear_subject_model}
\end{equation}

where $\boldsymbol{\theta}_s \in \mathbb{R}^{D+1}$ denotes a subject-specific parameter vector including a bias term. We refer to $\boldsymbol{\theta}_s$ as the \emph{subject preference embedding}, which serves as a compact representation of the perceptual characteristics of subject $s$.
Given the ratings collected from the training subjects, the subject preference embedding of each subject is first estimated using ridge regression:

\begin{equation}
\boldsymbol{\theta}_s
=
\arg\min_{\boldsymbol{\theta}}
\;
\|
\mathbf{r}_s
-
\mathbf{Z}\boldsymbol{\theta}
\|_2^2
+
\lambda_{\mathrm{ridge}}
\boldsymbol{\theta}^{\top}
\mathbf{R}
\boldsymbol{\theta},
\label{eq:ridge}
\end{equation}

where $\mathbf{Z}$ contains the augmented image representations $\mathbf{z}_i$, $\mathbf{r}_s$ contains the residual scores $r_{i,s}=Q_{i,s}-Q_i^{\mathrm{gen}}$ of subject $s$,
and $\mathbf{R}=\mathrm{diag}(1,\ldots,1,0)$ leaves the bias term unregularized.

To capture the overall distribution of subject preferences, we model the subject preference embeddings using a multivariate Gaussian distribution:
$
\boldsymbol{\theta}
\sim
\mathcal{N}
(
\boldsymbol{\mu},
\mathbf{\Sigma}
),
$
where the mean and covariance of the prior are estimated as

\begin{equation}
\boldsymbol{\mu}
=
\frac{1}{S}
\sum_{s=1}^{S}
\boldsymbol{\theta}_s,
\
\mathbf{\Sigma}
=
(1-\rho)
\mathbf{\Sigma}_{\mathrm{emp}}
+
\rho
\,\mathrm{diag}
\big(
\mathbf{\Sigma}_{\mathrm{emp}}
\big)
+
\epsilon \mathbf{I},
\label{eq:cov_shrink}
\end{equation}

where $\mathbf{\Sigma}_{\mathrm{emp}}$ is the empirical covariance of the training subject embeddings:
$
\mathbf{\Sigma}_{\mathrm{emp}}
=
\frac{1}{S-1}
\sum_{s=1}^{S}
\left(
\boldsymbol{\theta}_s-\boldsymbol{\mu}
\right)
\left(
\boldsymbol{\theta}_s-\boldsymbol{\mu}
\right)^{\mathrm{T}}.
$
The shrinkage parameter $\rho$ controls shrinkage toward a diagonal covariance, and $\epsilon$ is a small positive constant used for numerical stability. The prior is estimated only from the preference embeddings of the training subjects and remains fixed when adapting to a new subject.

For a previously unseen subject, only a small set of rated anchor images is available. 
Let $\mathcal{A}$ denote the anchor set, 
$\mathbf{r}_{\mathcal{A}}$ the corresponding residual scores, 
and $\mathbf{Z}_{\mathcal{A}}$ the associated image representations. 
The preference embedding is estimated by maximum a posteriori (MAP) adaptation:

\begin{equation}
\boldsymbol{\theta}^{*}
=
\arg\min_{\boldsymbol{\theta}}
\;
\|
\mathbf{r}_{\mathcal{A}}
-
\mathbf{Z}_{\mathcal{A}}
\boldsymbol{\theta}
\|_2^2
+
\lambda_{\mathrm{prior}}
(\boldsymbol{\theta}-\boldsymbol{\mu})^{\top}
\mathbf{\Sigma}^{-1}
(\boldsymbol{\theta}-\boldsymbol{\mu}).
\label{eq:map}
\end{equation}

The first term fits the observed anchor residuals, while the second regularizes the embedding toward the learned prior. 
When no anchor rating is available, the MAP estimate reduces to the prior mean,
$\boldsymbol{\theta}^{*}=\boldsymbol{\mu}$,
which provides the default preference profile before subject adaptation.
After adaptation, the estimated embedding $\boldsymbol{\theta}^{*}$ is used to predict the individualized quality scores of previously unrated images:
\begin{equation}
\hat{Q}_{i,s}
=
Q_i^{\mathrm{gen}}
+
\mathbf{z}_i^{\top}
\boldsymbol{\theta}^{*}.
\label{eq:personalized_prediction}
\end{equation}

\section{Experiments and Results}
\label{sec:experiments}

This section evaluates the proposed Generic-to-Individualized quality assessment
framework. We first introduce the construction and analysis of the GC360IQ database, including image acquisition, panorama generation, and subjective quality assessment. We then evaluate the proposed generic quality model through pseudo-data pretraining, performance comparison, and ablation studies. 
Finally, we evaluate subject-level quality prediction and analyze the learned subject preference space.

\begin{figure*}[t]
    \centering
    \includegraphics[width=0.85\textwidth,trim={0cm 0cm 0cm 0cm},clip]{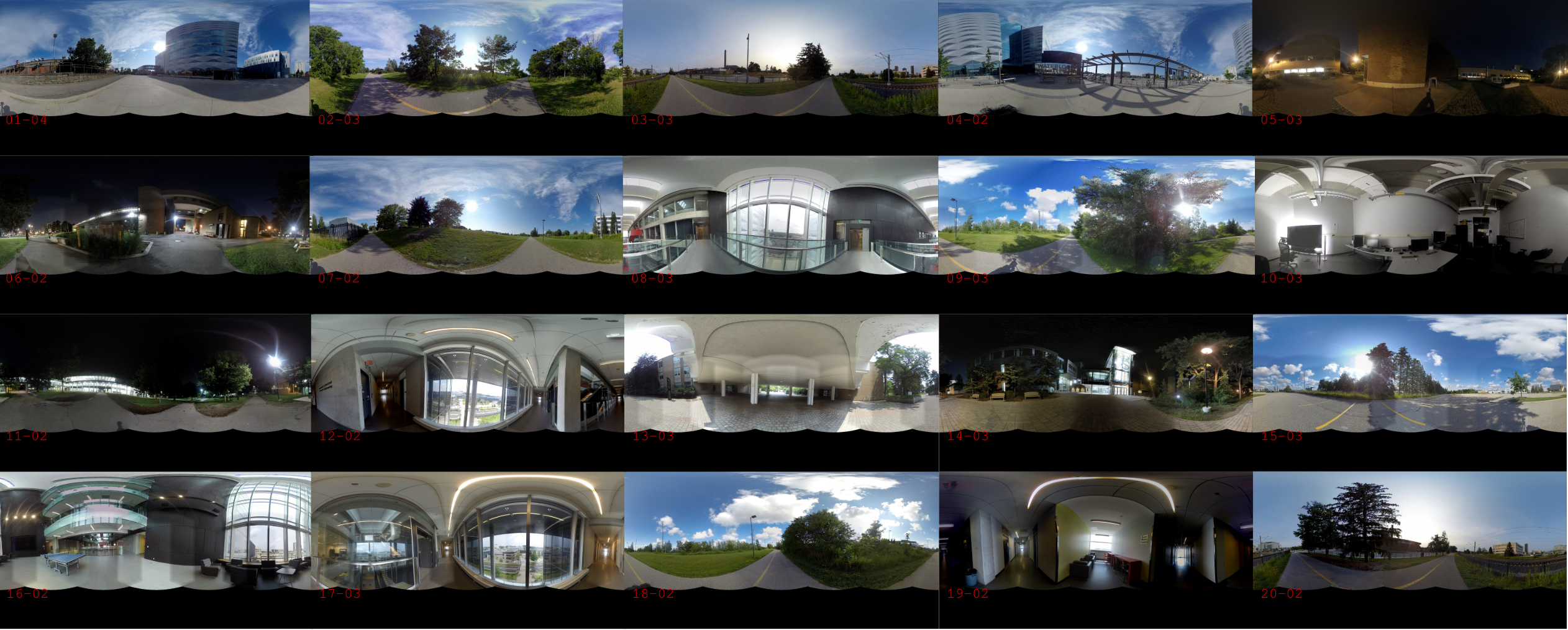}
        \caption{
        Overview of the 20 source scenes included in the GC360IQ database, covering outdoor daytime, nighttime, and indoor environments.
        }
    \label{fig:all_img}
\end{figure*}

\begin{figure}[t]
    \centering
    \includegraphics[width=0.65\linewidth]{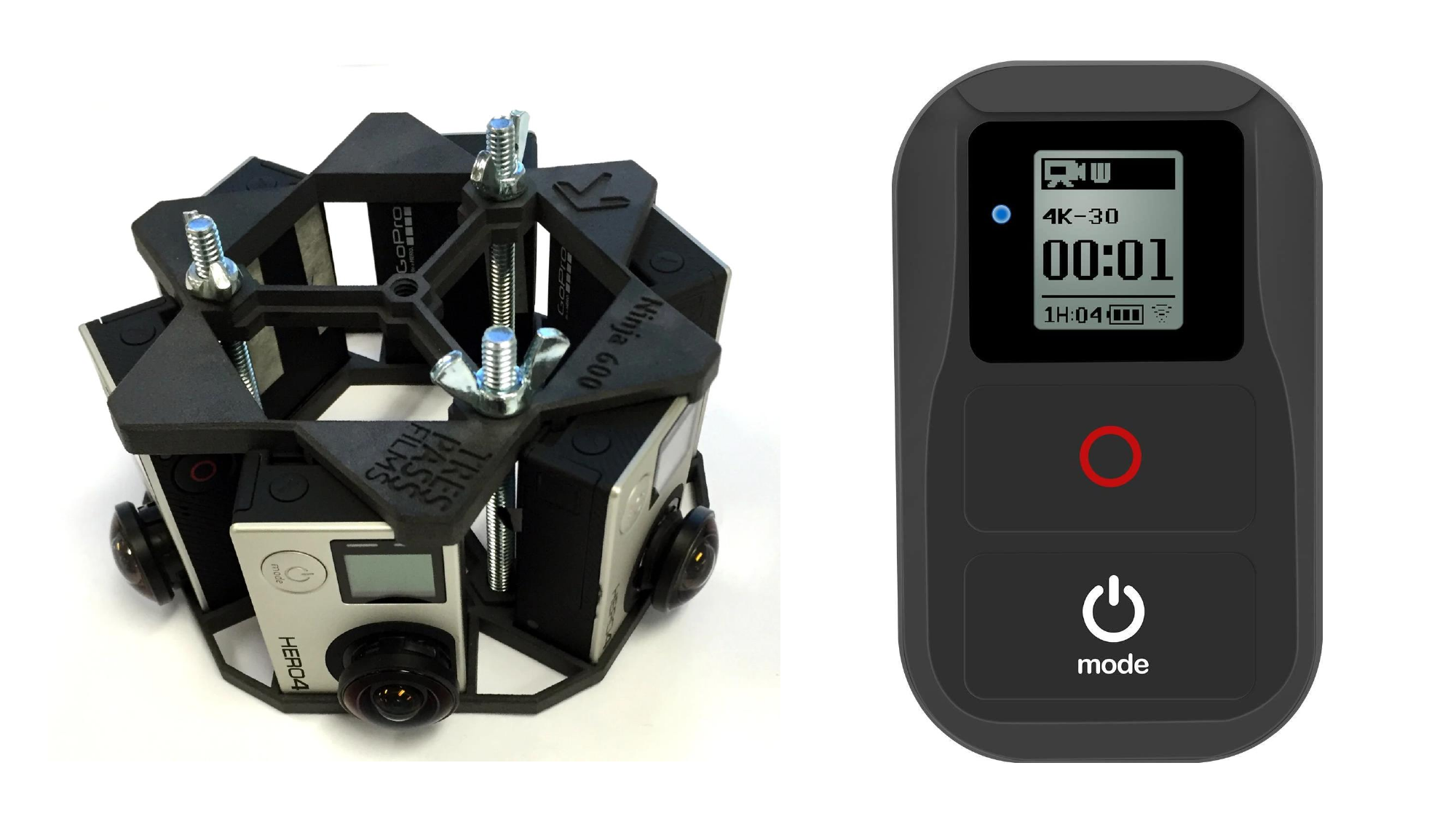}
    \caption{
    Seven-camera rig setup with GoPro Hero4 Black units. Six cameras capture equatorial views and one captures the zenith.
    }
    \label{fig:rig_setup}
\end{figure}

\subsection{Database Construction and Analysis}
\label{sec:database_constrc}

To support the development and evaluation of Generic-to-Individualized quality assessment for stitched 360-degree panoramas, we constructed the GC360IQ database through a complete pipeline including image acquisition, panorama generation, and subjective quality assessment. 
This subsection first describes the database construction process and then presents statistical analyses of the collected quality ratings, providing insights into both average perceptual quality and subject-dependent rating behaviors.

\subsubsection{Image Acquisition and Stitching}
\label{sec:image_acquisition_stitching}

A seven-camera GoPro Hero4 Black rig, shown in Fig.~\ref{fig:rig_setup}, was used to capture multi-view fisheye images for panoramic stitching. 
Six cameras were uniformly distributed along the equatorial plane ($\theta=90^\circ$, $\phi \in \{0^\circ,60^\circ,\dots,300^\circ\}$), while the remaining camera was oriented upward ($\theta=0^\circ$) to cover the zenith region. 
Automatic exposure was enabled for each camera so that every view used its direction-specific optimal exposure level. 
All cameras were triggered simultaneously via a wireless remote, ensuring precise temporal synchronization and sufficient overlap for stitching.

A total of 20 panoramic scenes were collected under diverse environmental conditions, including 10 outdoor daytime scenes, 4 nighttime scenes, and 6 indoor scenes. 
Figure~\ref{fig:all_img} provides an overview of the source scenes included in the database. 
The captured scenes cover a wide range of illumination levels, structural complexity, and visual content, providing diverse source material for the generation of stitched panoramas and subsequent quality assessment.

The captured multi-view fisheye images were first processed using PTGui Pro together with the panoramic alignment method~\cite{meng2020}. 
Geometric distortions and alignment errors were minimized during the stitching process to obtain well-aligned equirectangular (EQR) panoramas. 
The aligned panoramas were subsequently processed by different blending algorithms, allowing the effects of blending artifacts to be studied with minimal interference from geometric misalignment.

Each aligned EQR panorama was processed using five representative blending algorithms, including Naive Blending, Feather Blending (FB)~\cite{szeliski2006imageLinear}, Multi-Band Blending (MBB)~\cite{zhu2018comparative}, Mean-Value Coordinates Blending (MVCB)~\cite{farbman2011convolutionPyramid}, and Modified Poisson Blending (MPB)~\cite{tanaka2012seamlessPoisson}. 
The resulting panoramas contain two primary types of blending artifacts: luminance inconsistency, characterized by visible seams or brightness transitions across adjacent views, and detail loss, characterized by weakened object structures and texture details caused by large local brightness and contrast variations.

The 20 captured scenarios and five blending algorithms together form the GC360IQ database, which contains 100 stitched panoramic images in total.

\begin{figure*}[t]
    \centering
    \includegraphics[
        width=0.95\textwidth,
        trim={0cm 3.97cm 0cm 0cm},
        clip
    ]{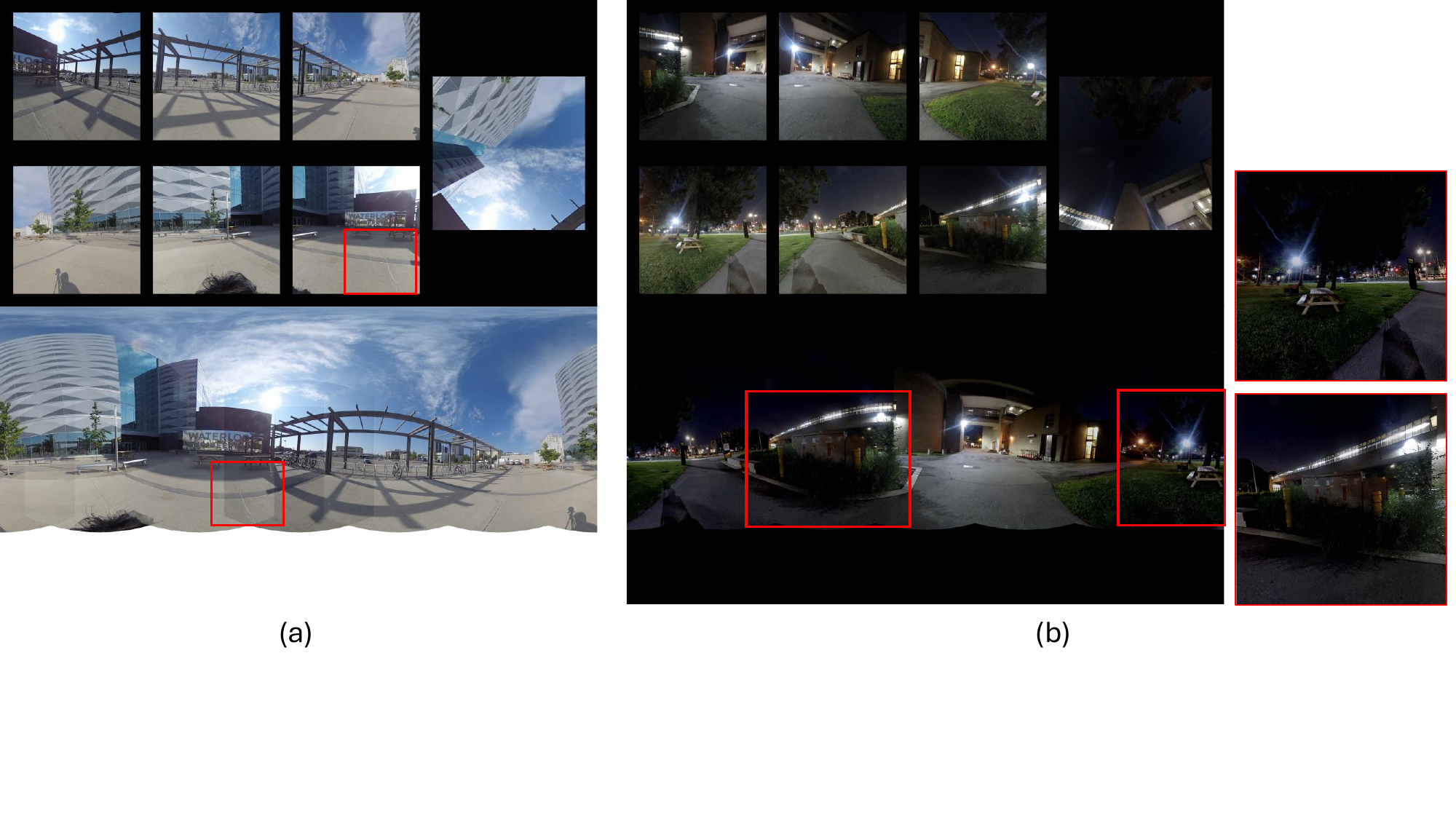}
    \caption{
    Representative examples of subject disagreement in stitched panorama quality assessment.
    In both examples, two groups of subjects give clearly different opinions on the same image, although the average score lies near the middle of the rating range.
    The red boxes indicate local stitching artifacts and the enlarged regions provide closer views of the corresponding distortions.
    In the left example, the highlighted region shows visible luminance inconsistency.
    In the right example, the highlighted regions show severe texture and detail loss in parts of the scene.
    These examples illustrate that MOS may hide meaningful subject-level differences in the perception of local stitching artifacts.
    }
    \label{fig:representative_disagreement_examples}
\end{figure*}

\subsubsection{Subjective Test}
\label{sec:subjective_test}

A controlled subjective test was conducted using a head-mounted display (HMD), 
allowing observers to freely explore each panorama in a full 360° environment while seated on a swivel chair. 
Thirty subjects (17 males and 13 females, aged 18--35) evaluated 100 stitched panoramas 
(20 scenarios $\times$ 5 blending algorithms). 
The order of scenarios and blending results was randomized to avoid ordering and fatigue bias.
All participants provided informed consent prior to the subjective experiment.

Each image was rated on three perceptual dimensions---\textit{luminance inconsistency}, 
\textit{detail loss}, and \textit{overall quality}---using a five-point discrete scale 
(1=\textit{Bad}, 5=\textit{Excellent}) following the Single-Stimulus ACR protocol~\cite{bt2002methodology}. 
Before scoring, subjects viewed two contextual visualizations inside the HMD: 
(1) a combined 2D FOV layout showing seven directional views extracted from the unblended inputs, 
and (2) a 360° seam visualization highlighting visible boundaries. 
Afterward, the five stitched results of each scenario were presented sequentially for rating relative 
to the internal visual expectation formed from the unblended views. 
Higher scores indicate smoother brightness across seams, better texture detail preservation, 
and higher overall perceptual quality. 
A short training session preceded the formal test. 
The obtained three scores per image were used as ground-truth labels for model training 
(Section~\ref{sec:gc360iq_train}).

All subjective scores $s_{i,j}$ were Z-score normalized per participant as 
$z_{i,j} = (s_{i,j} - \mu_j)/\sigma_j$ to remove rating bias, 
where $\mu_j$ and $\sigma_j$ are the mean and standard deviation of scores from subject $j$. 
Normalized scores were then rescaled to the original five-point range, 
and outliers were removed following ITU-R BT.500. 
The final mean opinion score (MOS) of image $i$ was obtained by averaging across subjects:
$
MOS_i = \frac{1}{N} \sum_{j=1}^{N} z_{i,j},
$
where $N$ is the number of valid participants.

\subsubsection{Subject Disagreement in Quality Assessment}
\label{sub_disagree}

After collecting subject-level ratings, we first examine whether MOS alone can adequately represent subjective quality for stitched panoramas. 
Although MOS is widely used as the ground truth in IQA, averaging subjective scores may overlook consistent differences among observers in their responses to the same stitching artifacts.

Figure~\ref{fig:representative_disagreement_examples} shows two representative examples. 
Both images have MOS values near the middle of the rating range, but their subject scores are clearly split between low and high ratings. 
For image~75, the MOS is 3.00, while 9 subjects assign a score of 2 and another 9 subjects assign a score of 4. 
For image~53, the MOS is 3.17, while 5 subjects assign a score of 2 and 10 subjects assign a score of 4; in addition, one subject assigns a score of 1 and another subject assigns a score of 5. 
These opposite opinions on the same images suggest that subjects may have different sensitivities to local stitching artifacts.

The highlighted regions further show that such disagreement is related to perceptually meaningful distortions. 
The red box in Fig.~\ref{fig:representative_disagreement_examples}(a) shows noticeable luminance inconsistency, while the enlarged regions in Fig.~\ref{fig:representative_disagreement_examples}(b) show severe texture and detail loss in parts of the scene. 
These examples motivate the modeling of subject-dependent quality preferences beyond MOS.

These observations provide the basis for the following analysis, where we examine whether subject-level deviations from the average rating can be represented by a structured subject preference space.

\subsection{Generic Quality Prediction}
\label{sec:generic_iqa_perf}

We evaluate the generic quality model on the proposed database and compare it with existing IQA methods. 
The effect of pseudo-data pretraining is also examined to verify its contribution to artifact-aware quality prediction.

\subsubsection{Pseudo Dataset Construction}
\label{sec:hdr_pseudo_data}

Since the GC360IQ database contains 100 stitched panoramas, training the generic quality model directly from scratch is limited by the database size. 
We therefore constructed a pseudo dataset from 250 publicly available HDR panoramas collected from Poly Haven~\cite{polyhaven}, including 125 outdoor scenes, 75 indoor scenes, and 50 nighttime scenes. 
This content distribution follows the same outdoor, indoor, and nighttime ratio as the GC360IQ database.

For each HDR panorama, seven virtual camera views were generated following the acquisition configuration described in Sec.~\ref{sec:image_acquisition_stitching}. 
To simulate view-dependent exposure differences, the views were tone mapped independently according to their local luminance statistics, and additional exposure offsets were introduced to create three exposure difference levels. 
The resulting multi-view images were then processed using the same stitching masks and blending pipeline as the GC360IQ database. 
The same five blending algorithms were applied to each exposure setting, producing $3\times5=15$ stitched panoramas for each HDR scene.

The pseudo dataset contains 3750 stitched panoramas in total and was used to pretrain the generic quality model before fine-tuning on the GC360IQ database with subjective quality ratings.
Rank-order pseudo labels were assigned to the generated samples to provide pretraining supervision rather than absolute MOS values.

\begin{table}[t]
\centering
\caption{Quantitative comparison of GC360IQ with existing IQA methods on the GC360IQ database.}
\renewcommand{\arraystretch}{1.15}
\scriptsize
\begin{tabular}{c l c c c}
\toprule
\textbf{Type} & \textbf{Methods} & \textbf{PLCC} & \textbf{SRCC} & \textbf{RMSE} \\
\midrule
\multirow{4}{*}{NR}
& NIQE~\cite{mittal2012makingNIQE}      & 0.0853 & 0.1065 & 0.8851 \\
& BRISQUE~\cite{mittal2012noBrisque}    & 0.0364 & 0.0406 & 0.8940 \\
& VGCN~\cite{xu2020blind}               & 0.4208 & 0.4282 & 0.8721 \\
& ST360IQ~\cite{tofighi2023st360iq}     & 0.6192 & 0.5889 & 0.7174 \\
\midrule
\multirow{8}{*}{FR}
& PSNR                                  & 0.0325 & 0.1167 & 0.8870 \\
& SSIM~\cite{wang2004imageSSIM}         & 0.3375 & 0.0962 & 0.8161 \\
& WS-PSNR~\cite{sun2017weightedWS-PSNR} & 0.0307 & 0.1023 & 0.8838 \\
& WS-SSIM~\cite{zhou2018weightedWS-SSIM}& 0.3208 & 0.0818 & 0.8310 \\
& FSIM~\cite{Zhang2011FSIM}             & 0.2807 & 0.2207 & 0.8277 \\
& VIF~\cite{Sheikh2006VIF}              & 0.6628 & 0.6734 & 0.6633 \\
& GMSD~\cite{xue2013gradient}           & 0.3238 & 0.3544 & 0.8433 \\
& DISTS~\cite{ding2020image}            & 0.6689 & 0.6801 & 0.6494 \\
\midrule
\multirow{1}{*}{--}
& \textbf{GC360IQ (Proposed)}   & \textbf{0.8986} & \textbf{0.8871} & \textbf{0.4092} \\
\bottomrule
\end{tabular}
\label{tab:ex_result}
\end{table}

\subsubsection{Training and Testing}
\label{sec:gc360iq_train}

All CNN models are trained and tested using scenario-wise splits to avoid content overlap between training and testing. 
For the pseudo dataset, HDR contents are divided into training and testing sets with an 80--20 split while preserving the proportions of outdoor daytime, nighttime, and indoor scenes. 
For the GC360IQ database, the 20 panoramic scenarios are divided into 10 training and 10 testing scenarios across different environment types, with all five blended panoramas from the same scenario kept in the same split. 
Each network predicts the MOS of its corresponding perceptual dimension, including \textit{luminance inconsistency}, \textit{detail loss}, and \textit{overall quality}.

The luminance-inconsistency and detail-loss branches are first pretrained on the pseudo dataset described in Sec.~\ref{sec:hdr_pseudo_data}. 
During fine-tuning on the GC360IQ database, all layers except the final fully connected regression layer are initially frozen. 
The final layer is optimized for 10 epochs using Adam with a learning rate of $1\times10^{-2}$ to align the pretrained output scale with the subjective scores. 
The entire network is then unfrozen and further optimized with a learning rate of $1\times10^{-3}$. 
The same two-stage fine-tuning strategy is applied to both artifact branches and the overall-quality prediction network.

After fine-tuning the two artifact branches, features from their first fully connected layers are extracted and concatenated as the fused quality representation. 
The overall-quality network is initialized from its pretrained counterpart and fine-tuned on the GC360IQ database using the same two-stage strategy as the artifact branches. 
It is trained on the fused representation to predict the MOS of overall quality.

All input tensors are normalized before training, and network weights are initialized using the default PyTorch initialization scheme unless inherited from pseudo-data pretraining. 
The Adam optimizer is used throughout training, and early stopping is applied when the validation loss does not improve for 10 consecutive epochs.

\begin{table}[t]
\centering
\caption{Ablation study on pseudo-data pretraining.}
\label{tab:pretrain_ablation}
\scriptsize
\begin{tabular}{llccc}
\toprule
\textbf{Target} & \textbf{Strategy}
& \textbf{PLCC}
& \textbf{SRCC}
& \textbf{RMSE} \\
\midrule

\multirow{3}{*}{\shortstack{Luminance\\Inconsistency}}
& Pretrain & 0.6851 & 0.7303 & 2.248 \\
& Scratch & 0.9208 & 0.8435 & 0.4841 \\
& Proposed & \textbf{0.9458} & \textbf{0.9004} & \textbf{0.4187} \\
\midrule

\multirow{3}{*}{\shortstack{Detail\\Loss}}
& Pretrain & 0.8298 & 0.6914 & 5.6141 \\
& Scratch & 0.9116 & 0.7651 & 0.4430 \\
& Proposed & \textbf{0.9249} & \textbf{0.7705} & \textbf{0.4001} \\
\midrule

\multirow{3}{*}{\shortstack{Overall\\Quality}}
& Pretrain & 0.6507 & 0.6519 & 0.8247 \\
& Scratch & 0.8555 & 0.8319 & 0.4378 \\
& Proposed & \textbf{0.8986} & \textbf{0.8871} & \textbf{0.4092} \\
\bottomrule
\end{tabular}
\end{table}

\subsubsection{Performance Comparison}
\label{sec:generic_results}

The generic quality model is evaluated on the GC360IQ database and compared with representative IQA methods. 
Three standard criteria are used: the Pearson Linear Correlation Coefficient (PLCC), the Spearman Rank-Order Correlation Coefficient (SRCC), and the Root Mean Square Error (RMSE).

Following common practice~\cite{duan2018perceptual}, a five-parameter logistic function is used to map the predicted scores to MOS values before PLCC and RMSE computation.
For non-learning methods, a random 50/50 split is used for logistic fitting and evaluation, and the process is repeated 50 times. 
Learning-based methods, including GC360IQ, VGCN, and ST360IQ, follow the scenario-wise split described in Sec.~\ref{sec:gc360iq_train}. 
Each model is trained and tested five times, and the average performance is reported. 
When model outputs are inversely correlated with MOS, the absolute SRCC value is used for consistency.

Table~\ref{tab:ex_result} presents the quantitative results on the GC360IQ database.
\textbf{GC360IQ} achieves the highest PLCC and SRCC and the lowest RMSE among all compared methods, showing the best agreement with subjective overall quality scores.
The results confirm the effectiveness of the proposed artifact-aware generic quality model for stitching-related distortion assessment.

\begin{figure}[t]
    \centering
    \includegraphics[width=\linewidth, trim={0.25cm 0.2cm 0.2cm 0.2cm}, clip]
    {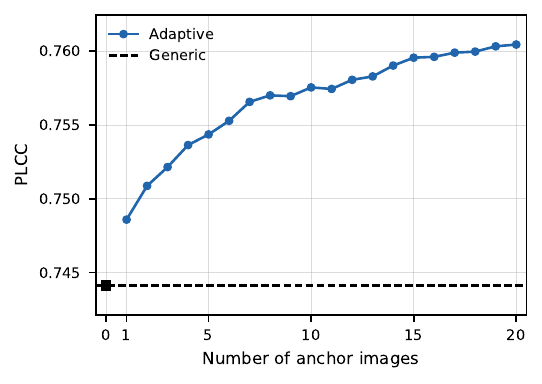}
    \caption{
    Subject-level prediction performance under random anchor order.
    The curves report the average results over 30 subjects and 15 random anchor orders.
    The generic prediction baseline is shown for reference.
    }
    \label{fig:random_anchor_rest80}
\end{figure}

\subsubsection{Pseudo-data Pretraining Ablation}
\label{sec:ablation}

Table~\ref{tab:pretrain_ablation} reports the effect of pseudo-data pretraining on luminance inconsistency, detail loss, and overall quality prediction. 
Three training strategies are compared: direct training on the GC360IQ database from scratch, direct evaluation of the pretrained model without fine-tuning, and the proposed strategy that fine-tunes the pretrained model on the GC360IQ database.

The PLCC and SRCC results show that useful stitching-related representations are learned before fine-tuning.
The relatively large RMSE values of direct pretraining are expected, since the pseudo labels provide rank-order supervision rather than absolute MOS values.

After fine-tuning on the GC360IQ database, the pretrained models outperform the scratch models for the luminance-inconsistency branch, the detail-loss branch, and overall quality prediction. 
This confirms the effectiveness of pseudo-data pretraining for learning artifact-aware quality representations.

\begin{table}[t]
\centering
\caption{Subject-level prediction performance on held-out images.}
\label{tab:heldout_final}
\scriptsize
\begin{tabular}{lccc}
\toprule
\textbf{Prediction}
& \textbf{PLCC}
& \textbf{SRCC}
& \textbf{RMSE} \\
\midrule
Generic prediction & 0.6169 & 0.6157 & 0.8236 \\
Subject-adaptive prediction & \textbf{0.6959} & \textbf{0.6809} & \textbf{0.7178} \\
\bottomrule
\end{tabular}
\end{table}

\begin{figure}[t]
    \centering
    \includegraphics[width=\linewidth, trim={0.25cm 0.25cm 0.2cm 0.2cm}, clip]
    {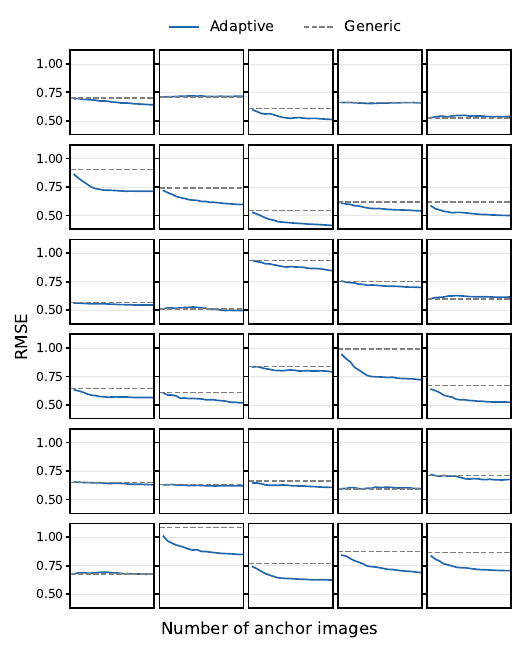}
    \caption{
    Subject-wise RMSE trajectories under random anchor order.
    Each panel shows the result of one subject.
    The adaptive result is averaged over 15 random anchor orders, and the dashed line denotes the generic prediction without subject adaptation.
    }
    \label{fig:personalized_rest80}
\end{figure}

\subsection{Subject-Adaptive Quality Prediction}
\label{sec:sub_adp_ana}

\subsubsection{Subject-Level Prediction Performance}
\label{sec:subject_level_prediction}

We first evaluate whether the proposed subject-adaptive model improves individual quality prediction beyond the generic quality estimate. 
Following the same scenario-wise split used for generic quality prediction, 50 images from the test scenarios are held out and are not used by either the generic model or the subject-adaptive model during training. 
For each split, the generic prediction is used as the baseline, while the subject-adaptive prediction is obtained by estimating the subject preference embedding from subject-specific ratings on the training images. 
The reported results are averaged over five repeated splits.

Table~\ref{tab:heldout_final} reports the results on held-out images. 
Compared with the generic prediction, the subject-adaptive model improves both PLCC and SRCC and reduces RMSE, showing that subject-level ratings provide useful information beyond the MOS-oriented generic predictor.

To further evaluate the adaptation behavior for a newly introduced subject, 
we use a repeated random-anchor protocol. 
The images are first divided into a fixed 20-image anchor pool and a fixed 80-image test set. 
The 80 test images are kept unchanged throughout the experiment, 
while the 20 anchor images are randomly ordered and progressively added as subject-specific observations. 
For a given anchor number $K$, the first $K$ images in the random order are used as anchors, 
assuming that these images have been rated by both the existing subjects and the new subject. 
The subject preference embedding is estimated from these anchor ratings, 
and prediction is performed on the fixed 80 test images of the same subject. 
Unlike the held-out image evaluation in Table~\ref{tab:heldout_final}, 
this setting focuses on predicting the unrated scores of a new subject from limited subject-specific observations.

Since all images in this experiment have been rated by the existing subjects, 
the generic quality estimate for each held-out subject is defined as the leave-one-subject-out MOS computed from the remaining subjects. 
Accordingly, the subject residual is measured relative to this estimate. 
When $K=0$, no rating from the held-out subject is available, and the preference
embedding reduces to the prior mean, $\boldsymbol{\theta}^{*}=\boldsymbol{\mu}$.
This $K=0$ embedding corresponds to the shared default preference profile and
produces the generic quality baseline. Because the residuals are centered with
respect to the leave-one-subject-out MOS, the prior mean is close to zero, and
the prediction reduces to the generic quality baseline.

The experiment is repeated 15 times with different random anchor orders, 
while the 20-image anchor pool and the 80-image test set remain fixed. 
In each repetition, all 30 subjects are treated as new subjects in turn, 
and the reported performance is averaged over subjects. 
As shown in Fig.~\ref{fig:random_anchor_rest80}, the PLCC consistently increases as more anchor ratings are provided, indicating improved agreement with the individual subject scores.
This result shows that the proposed model can better predict the quality scores of a new subject from a small number of rated images.

\begin{figure}[t]
    \centering
    \includegraphics[width=\linewidth, trim={0.25cm 0.2cm 0.2cm 0.2cm}, clip]{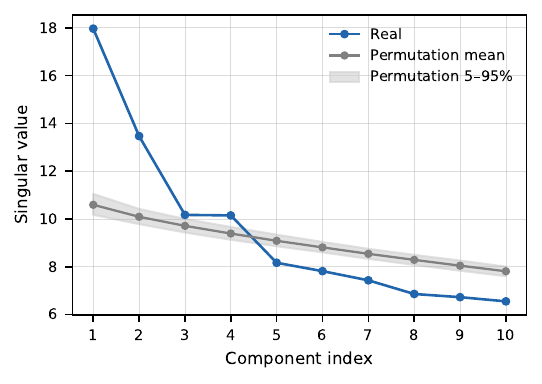}
    \caption{
    Singular value analysis of the subject residual matrix and the corresponding permutation baseline.
    }
    \label{fig:low_dim_struc}
\end{figure}

At the subject level, 
Fig.~\ref{fig:personalized_rest80} shows the RMSE curves for all 30 subjects under the same repeated random-anchor protocol. 
Each subject is shown in a separate panel, and each curve is averaged over the 15 random anchor orders. 
Although the average performance improves with increasing anchor numbers, 
the amount of improvement varies across subjects. 
For subjects whose ratings are already well approximated by the generic prediction, 
the baseline RMSE is relatively low and the additional gain from adaptation is limited. 
In contrast, subjects whose ratings deviate more from the average profile often have larger baseline errors, 
and their prediction performance is substantially improved after subject-specific adaptation. 
This indicates that the proposed Generic-to-Individualized model is especially useful for observers whose quality judgments are not well represented by the average prediction.


\subsubsection{Subject Preference Space}
\label{sec:subject_preference_space}

We next analyze the learned subject preference space. 
Specifically, we examine whether subject-level rating deviations form a compact structure and whether the learned embeddings are related to individual sensitivities to stitching artifacts.

We construct a subject residual matrix by removing the MOS from each subject's overall quality score. 
For image $i$ and subject $s$, the residual is defined as
\begin{equation}
r_{i,s}=Q_{i,s}^{(3)}-\mathrm{MOS}_i .
\end{equation}
This residual describes how the quality judgment of subject $s$ deviates from the average opinion for image $i$.

\begin{figure}[t]
    \centering
    \includegraphics[width=0.95\linewidth, trim={0.2cm 0.3cm 0.2cm 0.2cm}, clip]{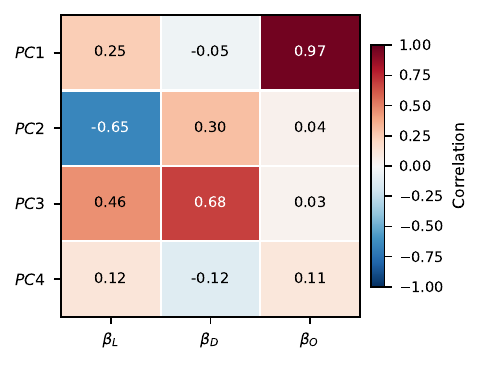}
    \caption{
    Correlation between the learned subject preference dimensions and independently estimated perceptual factors.
    }
    \label{fig:theta_beta_corr}
\end{figure}

We test this structure by comparing the singular values of the residual matrix with a shuffled baseline. 
The shuffled baseline removes subject-specific patterns while keeping the residual distribution unchanged. 
As shown in Fig.~\ref{fig:low_dim_struc}, 
the leading singular values of the original matrix are above the permutation mean, 
indicating that the subject-level deviations contain structure beyond random scoring fluctuations. 
This result supports the use of a learned preference space to model subject-dependent quality deviations.

Based on this observation, each subject is represented by a preference embedding $\boldsymbol{\theta}_s$ estimated by the subject-adaptive model. 
We use the embeddings obtained with $K=20$, where all anchor images are used for adaptation. 
For interpretation, the embeddings are projected onto the first four principal components of the learned preference space. 
This compact representation is used in the following analysis of subject-level artifact sensitivity.

\begin{figure*}[t]
    \centering
    \includegraphics[
        width=0.98\textwidth, trim={0.25cm 0.25cm 0.2cm 0.2cm}, clip
    ]{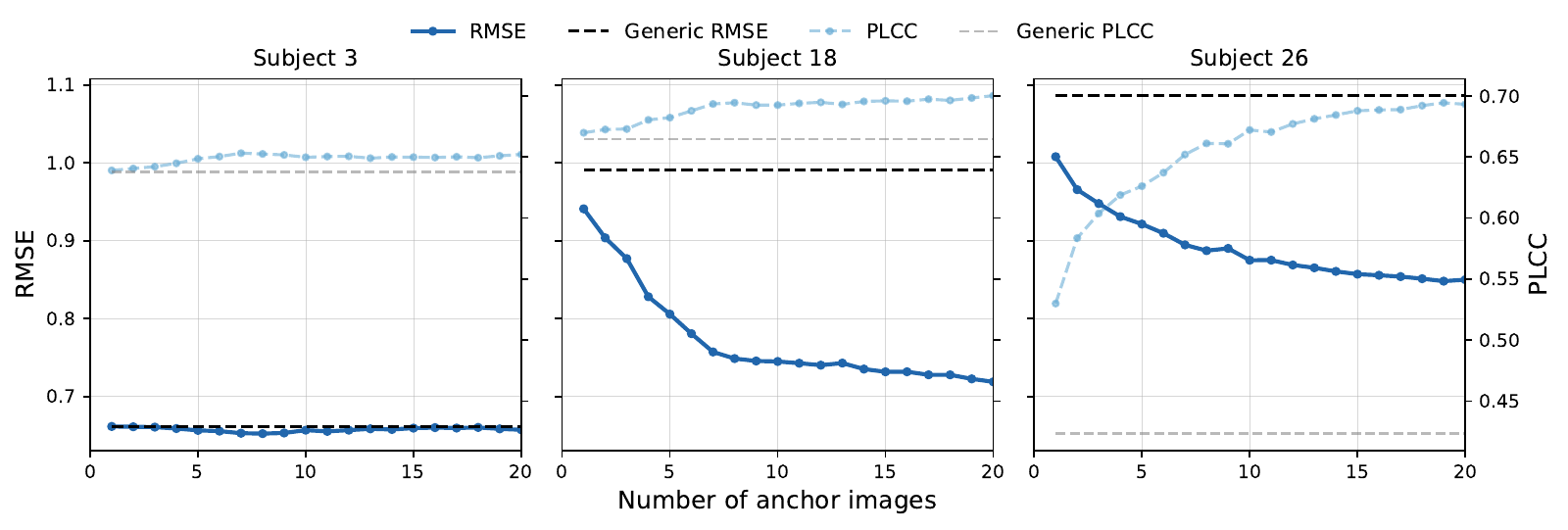}
    \caption{
    Representative subject-level prediction trajectories for three observers.
    The dark-blue solid curves and light-blue dashed curves denote the adaptive RMSE and PLCC, respectively, while the horizontal dashed lines indicate the corresponding generic baselines.
    }
    \label{fig:representative_subjects}
\end{figure*}

\begin{figure}[t]
    \centering
    \includegraphics[
        width=0.95\linewidth, trim={0.25cm 0.2cm 0.2cm 0.2cm}, clip
    ]{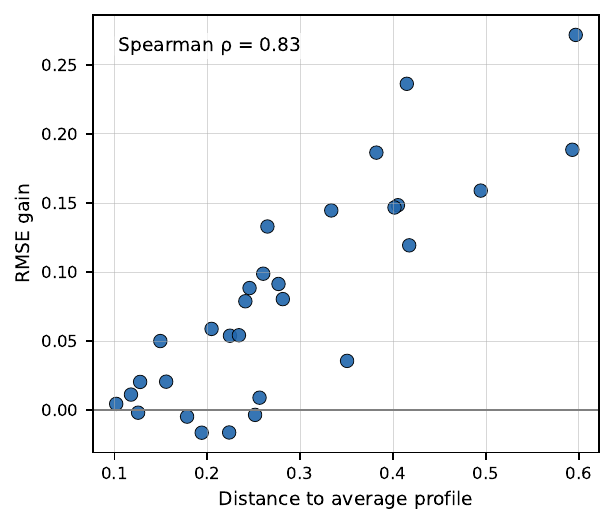}
    \caption{
    Relationship between the distance from the average preference profile and adaptation gain.
    Each point represents one subject.
    The horizontal axis denotes the distance from the average preference profile, and the vertical axis denotes the RMSE gain obtained by Generic-to-Individualized adaptation.
    }
    \label{fig:preference_distance_gain}
\end{figure}

\begin{figure}[t]
    \centering
    \includegraphics[
        width=\linewidth,
        trim={0cm 0.88cm 0cm 1.60cm},
        clip
    ]{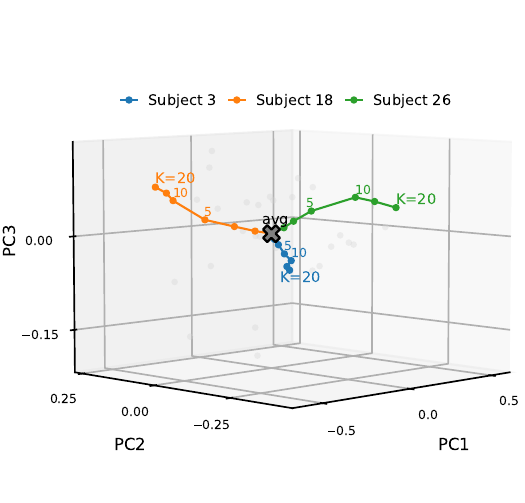}
    \caption{
    Preference adaptation trajectories for three representative subjects.
    The black cross denotes the shared average preference profile at $K=0$, and the trajectory endpoints correspond to the adapted embeddings with $K=20$ anchor images. The gray points show the $K=20$ embeddings of the remaining subjects.
    }
    \label{fig:preference_adaptation}
\end{figure}

To interpret these four dimensions, 
we estimate subject-level sensitivity parameters directly from the subjective scores. 
For each subject, the residual scores are fitted as
\begin{equation}
r_{i,s}
=
\beta_{O,s}
+
\beta_{L,s} a_L(i)
+
\beta_{D,s} a_D(i)
+
\epsilon_{i,s},
\end{equation}
where $a_L(i)$ and $a_D(i)$ denote the luminance-inconsistency and detail-loss artifact strengths of image $i$, respectively. 
The coefficients $\beta_{L,s}$ and $\beta_{D,s}$ measure the subject's sensitivity to the two artifact types, 
and $\beta_{O,s}$ measures the subject's overall scoring tendency relative to MOS.

Figure~\ref{fig:theta_beta_corr} compares the four preference dimensions with these fitted sensitivity parameters. 
The first dimension is strongly correlated with the overall scoring tendency $\beta_O$, while the second and third dimensions are more related to luminance and detail-loss sensitivities. 
This indicates that the learned preference space captures both subject-specific rating bias and artifact-dependent sensitivity.


\subsubsection{Preference Deviation and Adaptation Gain}
\label{sec:preference_recovery}

The learned preference space also explains why the benefit of subject adaptation varies across observers.
We measure each subject's preference deviation by the distance between its
$K=20$ preference embedding and the shared $K=0$ profile, which corresponds
to the prior mean $\boldsymbol{\mu}$ and the generic baseline position in the
preference space.
As shown in Fig.~\ref{fig:preference_distance_gain}, subjects farther from the shared $K=0$ profile tend to obtain larger RMSE gains from Generic-to-Individualized adaptation.
This indicates that subjects close to the average profile are largely captured by the generic predictor, 
whereas subjects with larger preference deviations benefit more from individualized adaptation.

At the subject level, Fig.~\ref{fig:personalized_rest80} shows the RMSE curves for all 30 subjects under the repeated random-anchor protocol.
Although the average performance improves with increasing anchor numbers, 
the amount of improvement varies across subjects. 
Subjects that are close to the average profile show relatively limited gains, 
whereas subjects with larger preference deviations obtain more noticeable improvements after adaptation.
We therefore select three representative subjects for a closer analysis of individualized prediction and preference recovery.

Subjects~3, 18, and 26 are selected as representative examples. 
Since the first three preference dimensions show the clearest correlations with the fitted perceptual factors analyzed in Sec.~\ref{sec:subject_preference_space},
Fig.~\ref{fig:preference_adaptation} visualizes their adaptation trajectories in the three-dimensional preference space.
The trajectories start from the shared preference profile at $K=0$, 
and the subsequent points correspond to the averaged embeddings obtained with $K=1,3,5,10,15,$ and $20$ anchor images under the repeated random-anchor protocol. 
Subject~3 remains close to the average profile as more anchor ratings are used, 
whereas Subjects~18 and~26 move away from the average profile toward different preference directions.

The corresponding prediction trajectories are shown in Fig.~\ref{fig:representative_subjects}. 
Subject~3 starts with relatively good prediction performance and only shows limited improvement after adaptation. 
In contrast, Subjects~18 and~26 have larger initial prediction errors and obtain clear gains as more anchor ratings are added. 
Subject~18 shows a strong RMSE reduction with a modest PLCC increase, 
while Subject~26 improves in both RMSE and PLCC. 
These examples show that the proposed method is particularly effective for subjects whose preferences deviate from the average profile.
By shifting the predictor from the generic quality estimate toward subject-specific preference regions, 
the model achieves the intended Generic-to-Individualized quality prediction and provides more accurate individualized assessment.

\section{Conclusion}

This paper presented GC360IQ, a Generic-to-Individualized IQA framework
that extends MOS-oriented stitching quality prediction to individualized
assessment. The GC360IQ database contains 100 stitched panoramas with
multidimensional quality ratings and complete individual scores from 30 subjects.
Using the unblended views as a perceptual reference, the generic model extracts
gradient and structural features around stitching regions to provide a reliable
baseline quality prediction. The individualized model constructs a compact
preference embedding space to represent each subject's deviation from this
baseline. By leveraging a learned preference prior and MAP adaptation, a new
subject is progressively mapped into the preference space as more anchor ratings
become available.

Experimental results show that the generic model outperforms representative IQA
methods and that subject adaptation improves individual quality prediction. The
subject residuals contain structured variation related to overall scoring tendency
and sensitivity to luminance inconsistency and detail loss, rather than merely
random rating noise. Subjects farther from the shared $K=0$ preference profile
also obtain larger gains from adaptation. These results demonstrate the value of
modeling individual ratings as a prediction target instead of treating observer
differences only as variation around MOS.

\bibliographystyle{IEEEtran}
\bibliography{references}

\end{document}